\documentclass[aps,pra,preprint,floatfix]{revtex4}
\usepackage{graphicx,amssymb,amsmath}
\usepackage{subfigure}
\newcommand{\eps}{\varepsilon}

\begin{document}
\title{Light Scattering on Two Parallel Subwavelength Cylinders}
\author{Alex Verevkin}
\affiliation{Institute of Automation and Electrometry, 
Siberian Branch, Russian Academy of Sciences,
1 Koptjug Avenue, Novosibirsk 630090, Russia}
\author{Alex Bereza}
\affiliation{Institute of Automation and Electrometry, 
	Siberian Branch, Russian Academy of Sciences,
	1 Koptjug Avenue, Novosibirsk 630090, Russia}
\author{David Shapiro}
\affiliation{Institute of Automation and Electrometry, 
Siberian Branch, Russian Academy of Sciences,
1 Koptjug Avenue, Novosibirsk 630090, Russia}

\begin{abstract}
	 The plasmon resonance has found important application in various systems, e.g., nanoantennas, solar panels, refractive index sensors. Unfortunately, a few analytical solutions for such systems are known. The work aims to find a solution for scattering by a plane electromagnetic wave on two parallel cylinders. When their diameters and the gap between them are less than the radiation wavelength, the quasistatic approximation is valid. We build up a conformal transformation that maps Cartesian into bipolar orthogonal coordinates and represent the scattered field as a decomposition by eigenfunctions of the Laplace operator. The near-field intensity distribution is shown to coincide with numerical calculation performed by COMSOL Multiphysics.  Green's function obtained allows one to find also higher decomposition orders. Compared to previous studies, we treat the general case of cylinders with different diameters and dielectric constants. 
\end{abstract}
\maketitle
\section{Introduction}

One of the most promising and exciting effects in modern nano-optics \cite{girard2005,shalaev2006} is a plasmon resonance, the excitation of electron plasma oscillations in a metal under the electromagnetic waves. These oscillations lead to the local field enhancement when the incident field could increase by several orders \cite{stockman2011,novotny2012}. Moreover, the light localization at the nanoscale can be achieved by an optical resonance in pure dielectric structures with no metallic elements \cite{devilez2015,krasnok2018}.  Therefore, it is possible to obtain a considerable intensity in a particular geometrical structure.

The applications of plasmon resonances are various. For example, plasmonic structures in the photovoltaic panel can significantly enhance its efficiency by increasing the energy conversion ratio of the incident wave. Usually, one increases the intensity by confining the semiconductors' energy, where plasmon structures create additional electron-hole pairs \cite{enrichi2018}. The consequent application is the optical sensors of the refraction index based on the plasmon resonance effect \cite{homola2008,shalabney2011,piliarik2012}. By analyzing the spectrum of reflection and shift of the resonance peaks in the angular dependence, one can reach high values of the resolution and sensitivity \cite{xu2019} in conjunction with their comparatively small size and the possibility of repeated use. For instance, one improves Bragg sensors covering the fiber surface by nanoparticles \cite{bialiayeu2012,arasu2016}. However, all these applications face a significant obstacle. Due to the non-trivial geometry, it is almost impossible to find analytical solutions for a scattered amplitude. Current methods for this problem analysis are numerical: Finite-Difference Time-Domain method (FDTD) \cite{yee1966numerical}, Discrete Dipole Approximation (DDA) \cite{purcell73,Draine08}, Boundary Element Method (BEM) \cite{brebbia2016boundary}, Finite Element Method (FEM) \cite{zienkiewicz2005finite}. However, each method has limitations and numerical errors, which can not be quickly evaluated and determined. Therefore, new analytical solutions must assist in choosing an optimal configuration for the plasmon structure. At least it is helpful to verify the numerical simulations. 

The present paper study the scattering of plane waves by two parallel cylinders. The problem is two-dimensional, and the aim is to find the field distribution using a conformal transformation. Such a quasistatic approach is valid for subwavelength particles \cite{VKS7}. A conformal transformation for two cylinders, the bipolar coordinates, had been found earlier for the symmetric situation of equal cylinders \cite{vorobev2010}. Below, we expand the method to cylinders with different diameters and dielectric constants. Ref. \cite{lei2010} has considered the system of two wires for a specific case, when their cross-sections have a mutual point. The authors call it the "kissing nanowires." Energy concentration and confinement occur in the singularity, a point of contact. Surface plasmon modes propagate toward the singularity, place of the energy accumulation. There is an alternative way to obtain analytical approximate solutions, the Born approximation modified for dielectric particles \cite{bereza2017,bereza2019,PhysRevRes2020,PhysRevA.104.063514,Ustimenko2022}. Chains of parallel infinite cylinders also permits approximate analytical solutions in the form of decomposition over cylindrical waves \cite{belan2015,lee2016}.

We introduce the geometry in Sect. \ref{sec:statement} and discuss the conformal transformation in Sect. \ref{sec:transformation}. Next, in Sect. \ref{sec:quasistatic}, we find the zero-order solution in the form of decomposition by eigenfunctions. The near-field intensity distribution is compared with numeical calculation.  Sect. \ref{sec:conclusions} summarizes the conclusions. We present Green's function, allowing one to calculate the next orders, in Appendix. 

\section{Laplace equation}\label{sec:statement}

\begin{figure}
\includegraphics[width=0.5\textwidth]{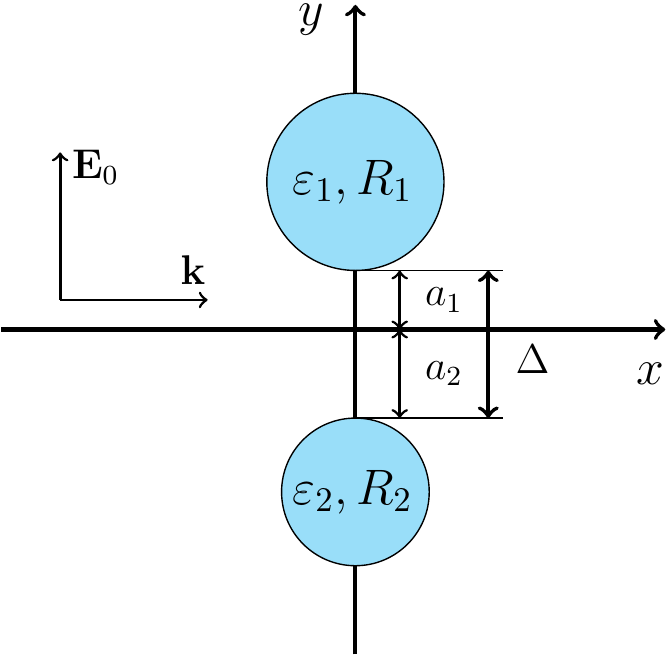}
\caption{Sketch of geometry.}
\label{fig:sketch}
\end{figure}

Fig. \ref{fig:sketch} shows the cross-section of considered geometry. Parallel cylinders of radius $R_{1}$ and $R_{2}$, with dielectric permittivity $\varepsilon_{1}$ and $\varepsilon_{2}$, respectively, interact with plane electromagnetic wave of amplitude $\vec{E}_{0}$ and wavevector $\vec{k}$. A gap between two cylinders has width $\Delta$. Axis $y$ of the system goes through centers of circles, while $z$ is parallel to the cylinder axes. Validity condition is that the wavelength of incident wave $\lambda$ should be greater than each characteristic dimension of our system 
\begin{equation}\label{validity}
kb\ll1,
\end{equation}
where $b=\max(R_1,R_2,\Delta)$. The condition is necessary to employ the static approach.

Consider the equation for potential $\phi$ under wave of $p$-polarization:
\begin{equation}
\Delta\phi+k^{2}\phi = 0.
\end{equation}
Then one can decompose potential $\phi$ into a series that includes only even orders:
\begin{equation}\label{decomposition}
\phi(x) = \phi^{(0)}(x) + k^{2}\phi^{(2)}(x)+\dots
\end{equation}

We get a chain of equations
\begin{align}
\Delta\phi^{(0)}(x) = 0,
\label{laplace:eq}\\
\Delta\phi^{(n)}(x) = -\phi^{(n-2)}(x),
\label{pois:eq}
\end{align}
where $n\geqslant1$.

Boundary conditions are
\begin{align}
\left.\varepsilon_{in}\frac{\partial \phi_{in}}{\partial {n}}\right|_{\gamma} = \left.\frac{\partial \phi_{out}}{\partial {n}}\right|_{\gamma},
\left.\phi_{in}\right|_{\gamma} = \left.\phi_{out}\right|_{\gamma},
\label{borderin2:eq}
\end{align}
where $\phi_{in}$ is the potential inside the cylinder, $\phi_{out}$ is the potential outside,  $\varepsilon_{in}$ is the dielectric permittivity, $\gamma$ is the boundary of the cylinder, $\vec{n}$ is the unit vector normal to the border.
So then, one has to solve the Laplace equation  (\ref{laplace:eq}) in the zero-order. Then, for the next orders we deal with Poisson equation (\ref{pois:eq}). One can find all next orders applying Green's function for the Poisson equation derived in the Appendix.

These conditions will apply to all orders, except of the zero-order, where one has to include the incident field. Therefore, zero-order boundary conditions are:
\begin{align}
\left.\varepsilon_{in}\frac{\partial \phi^{(0)}_{in}}{\partial \vec{n}}\right|_{\gamma} = \left.\frac{\partial \phi^{(0)}_{out}}{\partial \vec{n}}\right|_{\gamma}+\left.\frac{\partial \phi_{0}}{\partial n}\right|_{\gamma},\quad
\left.\phi^{(0)}_{in}\right|_{\gamma} = \left.\phi^{(0)}_{out}\right|_{\gamma}+\left.\phi_{0}\right|_{\gamma},
\label{conditions_0-order}
\end{align}
where $\phi_{0}$ is the potential of incident field. In quasistatic approximation the field is uniform, and then its potential is a linear coordinate function
\begin{equation}
\phi_{0} = -E_{0}y.\label{fall_field:eq}
\end{equation}

\section{Conformal transformation}\label{sec:transformation}

After a conformal mapping, a harmonic function remains harmonic \cite{ablowitz2003,brown2009}. Then we look for a conformal transformation to build up an analytical formula. To construct the map, we must analyze its level lines that must walk along boundaries.
 
The appropriate map (bipolar coordinates) is described by the following formula \cite{vorobev2010}:
\begin{equation}\label{bipolar}
w(z) = \xi+i\eta=\ln\left(\frac{z+iC}{z-iC}\right),
\end{equation}
where $C$ is a real constant value. The transformation can be also rewritten as
\begin{equation}
e^{\xi}e^{i\eta} = \frac{x+iy+iC}{x+iy-iC}.
\end{equation}
Direct and inverse transformations are
\begin{align}
x = \frac{C\sin\eta}{\cosh\xi-\cos\eta},
&\quad&
y = \frac{C\sinh\xi}{\cosh\xi-\cos\eta}.
\nonumber
\\
\tanh\xi = \frac{2Cy}{x^2+y^2+C^2},
&\quad&
\tan\eta = \frac{2Cx}{x^2+y^2+C^2}.
\label{coord2:eq}
\end{align}
Excluding $\xi,\eta$ we can obtain parametric equations for level lines:
\begin{eqnarray}
	x^2+(y-C\coth\xi_0)^2=\frac{C^2}{\sinh^2\xi_0},
	\quad\xi_0\neq0,
\nonumber
\\
	(x+C\cot\eta_0)^2+y^2=\frac{C^2}{\sin^2\eta_0},
	\quad	\eta_0\neq0.
\end{eqnarray}

\begin{figure}
\begin{center}
\includegraphics[width=0.62\textwidth]{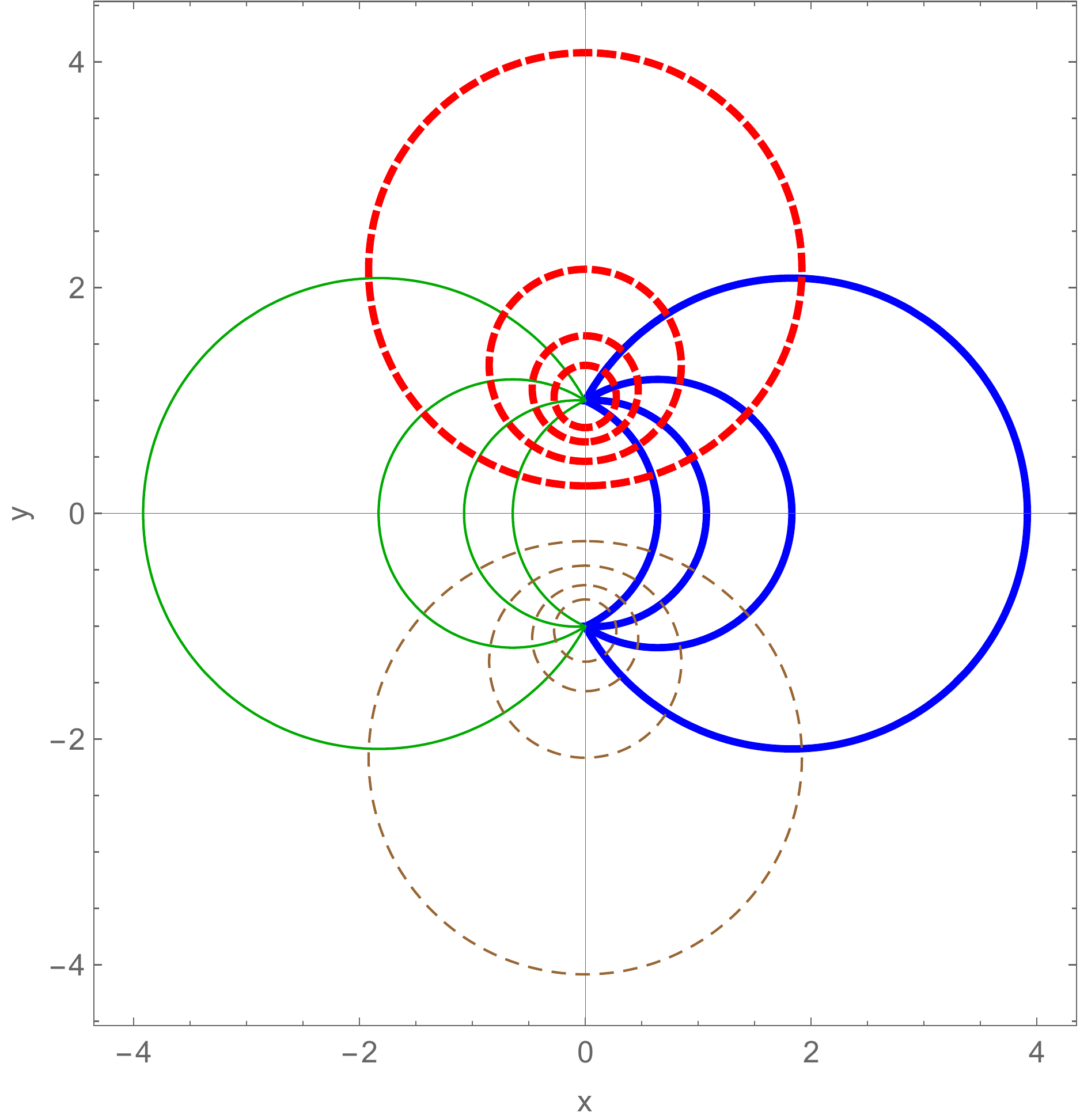}
\caption{Level lines (\ref{coord2:eq}) of the transformation in $(x,y)$ plane at $C=1$: $\xi=0.5,1,1.5,2,\eta\in(-\pi,\pi)$ (solid thick lines); $\xi=-0.5,-1,-1.5,-2,\eta\in(-\pi,\pi)$ (solid thin); $\eta=0.5,1,1.5,2,\xi\in(-\pi,\pi)$ (dashed thick); $\eta=-0.5,-1,-1.5,-2,\xi\in(-\pi,\pi)$ (dashed thin).}
\label{fig:bipolar}
\end{center}
\end{figure}

As one can see the first family of level lines are circles with radius $|C/\sinh\xi_0|$ shifted to $\pm C\coth\xi_0$ in top and bottom semi-planes. Similarly  the second family consists of circles with radius $|C/\sin\eta_0|$ shifted to $\pm C\cot\eta_0$ in left and right, as shown in Fig. \ref{fig:bipolar}. Constant $C$ defines the coordinate origin and represents the equipotential values for the borders. It also allows working with asymmetric case, if we use top and bottom dashed circles of different radius. These formulas define the origin of coordinate system and determine the boundary values in new coordinates, when circle are mapped to constant $\xi$, with $\eta \in (-\pi,\pi)$. The level lines are circles or their segments. The transformation properly maps the boundaries of top and bottom circles into equipotential lines. 

Laplace's operator in the $\eta$, $\xi$ coordinates is
\begin{equation}
	\Delta_{\xi,\eta} = \frac{(\cosh\xi-\cos\eta)^{2}}{C^{2}}\left(\frac{\partial^{2}}{\partial\xi^{2}}+\frac{\partial^{2}}{\partial\eta^{2}}\right).
\end{equation}

For circles $R_1,R_2$ and gap dimension $\Delta$, one obtain the following equations
\begin{align}
	\frac{C}{\tanh\xi_{1}}=R_{1}+a_{1},\quad
	\frac{C}{\sinh\xi_{1}}=R_{1},\nonumber
	\\
	\frac{C}{\tanh\xi_{2}}=-R_{2}-a_{2},\quad
	\frac{C}{\sinh\xi_{2}}=-R_{2},\nonumber
	\\
	a_{1}+a_{2} = \Delta,
	\label{border}
\end{align}
where $\xi_{1}, \xi_{2}$ are equipotential values for the boundaries. One can see that Eq. (\ref{border}) is symmetric with respect to transformation $y\to-y$: $R_1\leftrightarrow-R_2, a_1\leftrightarrow-a_2,\Delta\to-\Delta$.  

\begin{figure}
	\includegraphics[width=4in]{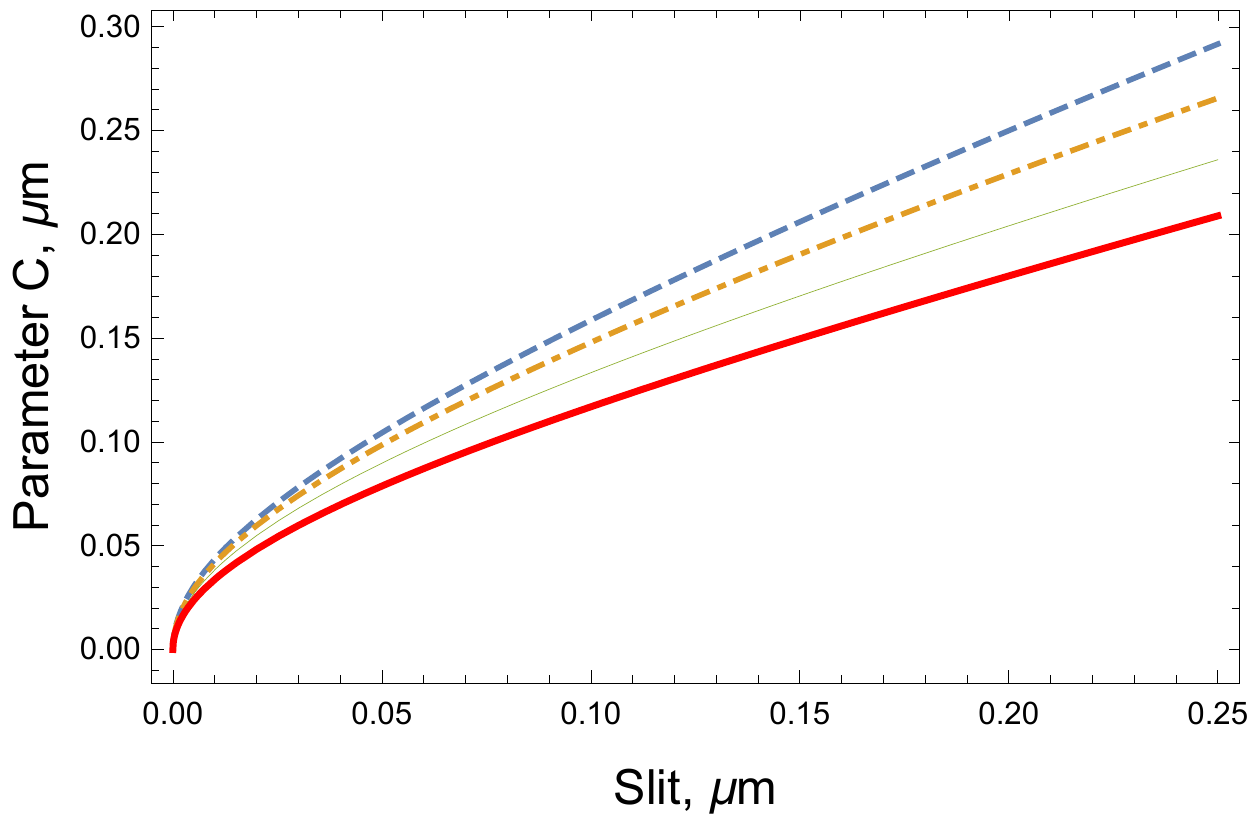}
\caption{Constant $C$ as a function of slit dimension $\Delta$ at fixed $R_1=0.1~\mu$m and different $R_2=1,$ (dashed line), 0.5 (dot-dashed), 0.25 (thin), 0.125 (thick)}	
\label{fig:slit}
\end{figure}

Solving equations we get
\begin{align}
	\xi_{1} = \ln\left(1+\frac{a_{1}}{R_{1}}+\sqrt{\left(1+\frac{a_{1}}{R_{1}}\right)^{2}-1}\right),\nonumber\\
	\xi_{2} = \ln\left(1+\frac{a_{2}}{R_{2}}-\sqrt{\left(1+\frac{a_{2}}{R_{2}}\right)^{2}-1}\right),\nonumber\\
	a_{1} = \frac{\Delta(\Delta+2R_{2})}{2(\Delta+R_{1}+R_{2})},
	\quad
	a_{2} = \frac{\Delta(\Delta+2R_{1})}{2(\Delta+R_{1}+R_{2})},\nonumber\\C =\frac{\sqrt{\Delta(\Delta+2R_1)(\Delta+2R_2)(\Delta+2R_1+2R_2)}}{2(\Delta+R_1+R_2)}.\label{constants}
\end{align}
In limiting cases of a narrow or wide slit the following asymptotics are valid
\begin{equation}\label{limits}
C\approx\begin{cases}
\sqrt{\frac{2\Delta R_1R_2}{R_1+R_2}}+O(\Delta^{3/2}),&\Delta\ll R_{1,2},\\
\frac12(\Delta+R_1+R_2)+O(\Delta^{-1}),&\Delta\gg R_{1,2}.	
\end{cases}	
\end{equation}
The constant $C$ is plotted in Fig.~\ref{fig:slit} as a function of slit dimension $\Delta$. 
We see the characteristic behaviour (\ref{limits}): at  small $\Delta\ll R_{1,2}$ the square root dependence takes place $C\sim\Delta^{1/2}$; at large $\Delta\gg R_{1,2}$
it becomes linear $C\sim\Delta$.

For two equal cylinders $R_1=R_2=R,\eps_1=\eps_2, a_1=a_2=a$  Eq. (\ref{constants}) yields $a=\Delta/2,\xi_1=-\xi_2,C\approx\sqrt{aR}$ ($a\ll R$). The solution turns into symmetric problem studied by Vorob'ov \cite{vorobev2010}. At $\Delta,C\to0$, in the limit of "kissing cylinders", Eq. (\ref{bipolar}) reduces to an inversion transformation introduced by Lei et al \cite{lei2010}:
\[
w(z)\approx\frac{2iC}{z}+O(C^3).
\] 
Let us return to the general asymmetric case.

Boundary conditions (\ref{conditions_0-order}) in bipolar coordinates take into account two borders:
\begin{align}
\left.\phi^{(0)}_{in}(x)\right|_{\xi_{1}} = \left.\phi^{(0)}_{out}(x)\right|_{\xi_{1}}+\left.\phi_{0}(x)\right|_{\xi_{1}},\nonumber\\
\left.\phi^{(0)}_{in}(x)\right|_{\xi_{2}} = \left.\phi^{(0)}_{out}(x)\right|_{\xi_{2}}+\left.\phi_{0}(x)\right|_{\xi_{2}},\nonumber\\
\varepsilon_{1}\left.\frac{\partial\phi^{(0)}_{in}(x)}{\partial\xi}\right|_{\xi_{1}} = \left.\frac{\partial\phi^{(0)}_{out}(x)}{\partial\xi}\right|_{\xi_{1}}+\left.\frac{\partial\phi_{0}(x)}{\partial\xi}\right|_{\xi_{1}},\nonumber\\
\varepsilon_{2}\left.\frac{\partial\phi^{(0)}_{in}(x)}{\partial\xi}\right|_{\xi_{2}} = \left.\frac{\partial\phi^{(0)}_{out}(x)}{\partial\xi}\right|_{\xi_{2}}+\left.\frac{\partial\phi_{0}(x)}{\partial\xi}\right|_{\xi_{2}}.
\label{border4:eq}
\end{align}

\section{Quasistatic approximation}\label{sec:quasistatic}

It is easy to find eigen functions of the Laplace operator in bipolar coordinates
\begin{equation}
 \psi_m(\xi,\eta)= e^{\pm m\xi}\cos(m\eta).   
\end{equation}
Then, we can determine the zero order as a series of the eigen functions: 
\begin{equation}
\phi^{(0)}=
\begin{cases}
\sum_{m=0}^{+\infty}A^{(0)}_{m}e^{-m\xi}\cos(m\eta), &\xi>\xi_{1},
\\
\sum_{m=0}^{+\infty}\left(C^{(0)}_{m}e^{m\xi}+D^{(0)}_{m}e^{-m\xi}\right)\cos(m\eta),&\xi_{2}<\xi<\xi_{1},
\\
\sum_{m=0}^{+\infty}B^{(0)}_{m}e^{m\xi}\cos(m\eta), &\xi<\xi_{2}.
\label{pot:eq}
\end{cases}
\end{equation}

Area inside the first/second cylinder corresponds to $\xi>\xi_{1}$/$\xi<\xi_{2}$, respectively. The interval $\xi_2<\xi<\xi_1$ is an external space. After solving boundary conditions equations at borders $\xi=\xi_1,\xi_2$ we obtain
\begin{align}
A^{(0)}_{m}=\frac{-4E_{0}Ce^{2m\xi_{1}}\left(e^{2m\xi_{2}}(\varepsilon_{2}-1)+(\varepsilon_{2}+1)\right)}{J_{m}},\nonumber\\
B^{(0)}_{m}=\frac{4E_{0}C\left(e^{2m\xi_{1}}(\varepsilon_{1}+1)+(\varepsilon_{1}-1)\right)}{J_{m}},\nonumber\\
C^{(0)}_{m}=\frac{2E_{0}C(\varepsilon_{1}-1)\left(e^{2m\xi_{2}}(\varepsilon_{2}-1)+(\varepsilon_{2}+1)\right)}{J_{m}},\nonumber\\
D^{(0)}_{m}=\frac{-2E_{0}Ce^{2m\xi_{2}}(\varepsilon_{2}-1)\left(e^{2m\xi_{1}}(\varepsilon_{1}+1)+
(\varepsilon_{1}-1)\right)}{J_{m}}.\nonumber\\
J_{m} = e^{2m\xi_{1}}(\varepsilon_{1}+1)(\varepsilon_{2}+1)-e^{2m\xi_{2}}(\varepsilon_{1}-1)(\varepsilon_{2}-1),\quad m\neq0.
\end{align}

\begin{figure}
\subfigure[]%
{\includegraphics[width=.27\textwidth]{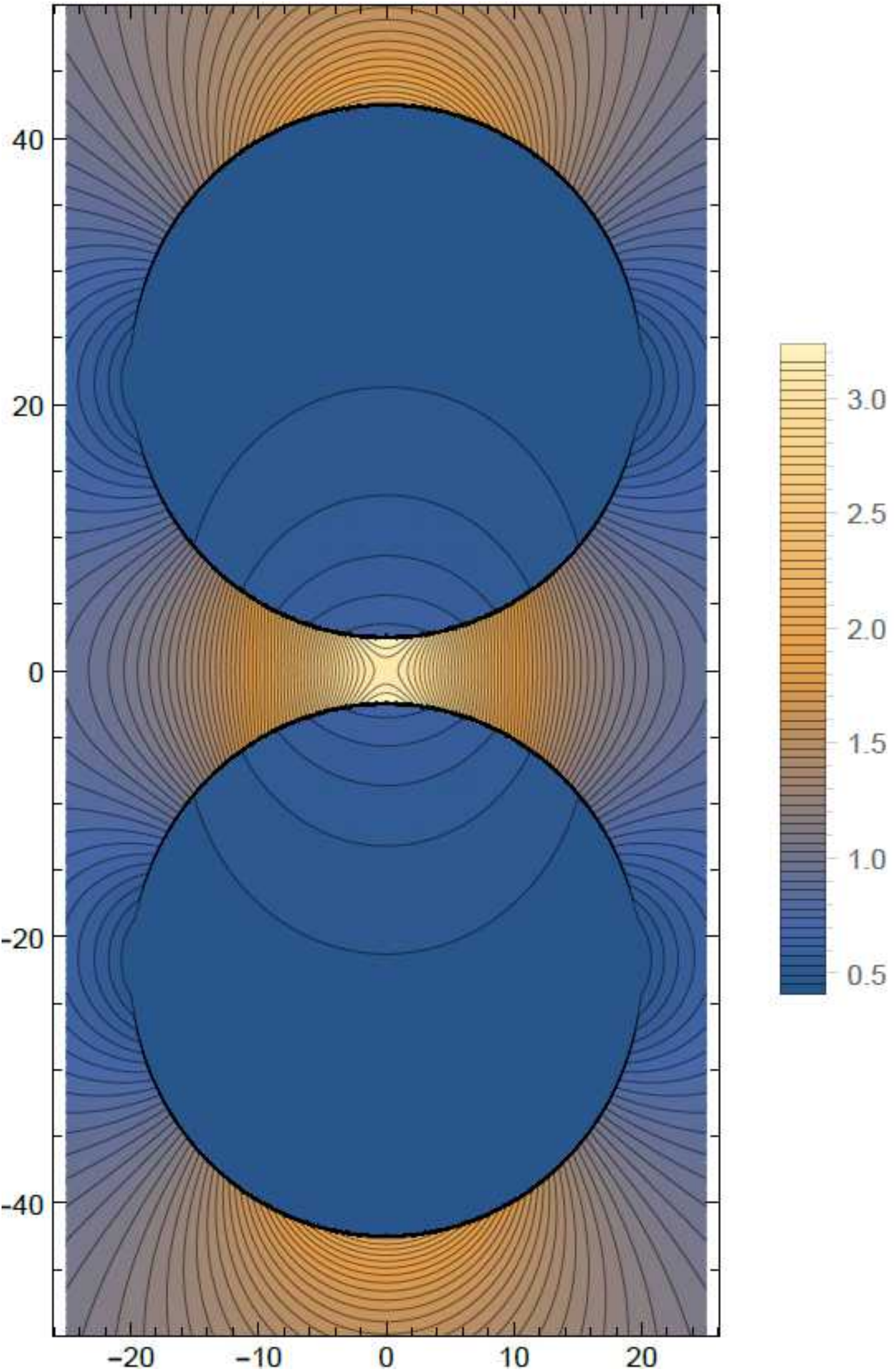}}
\subfigure[]%
{\includegraphics[width=.55\textwidth]{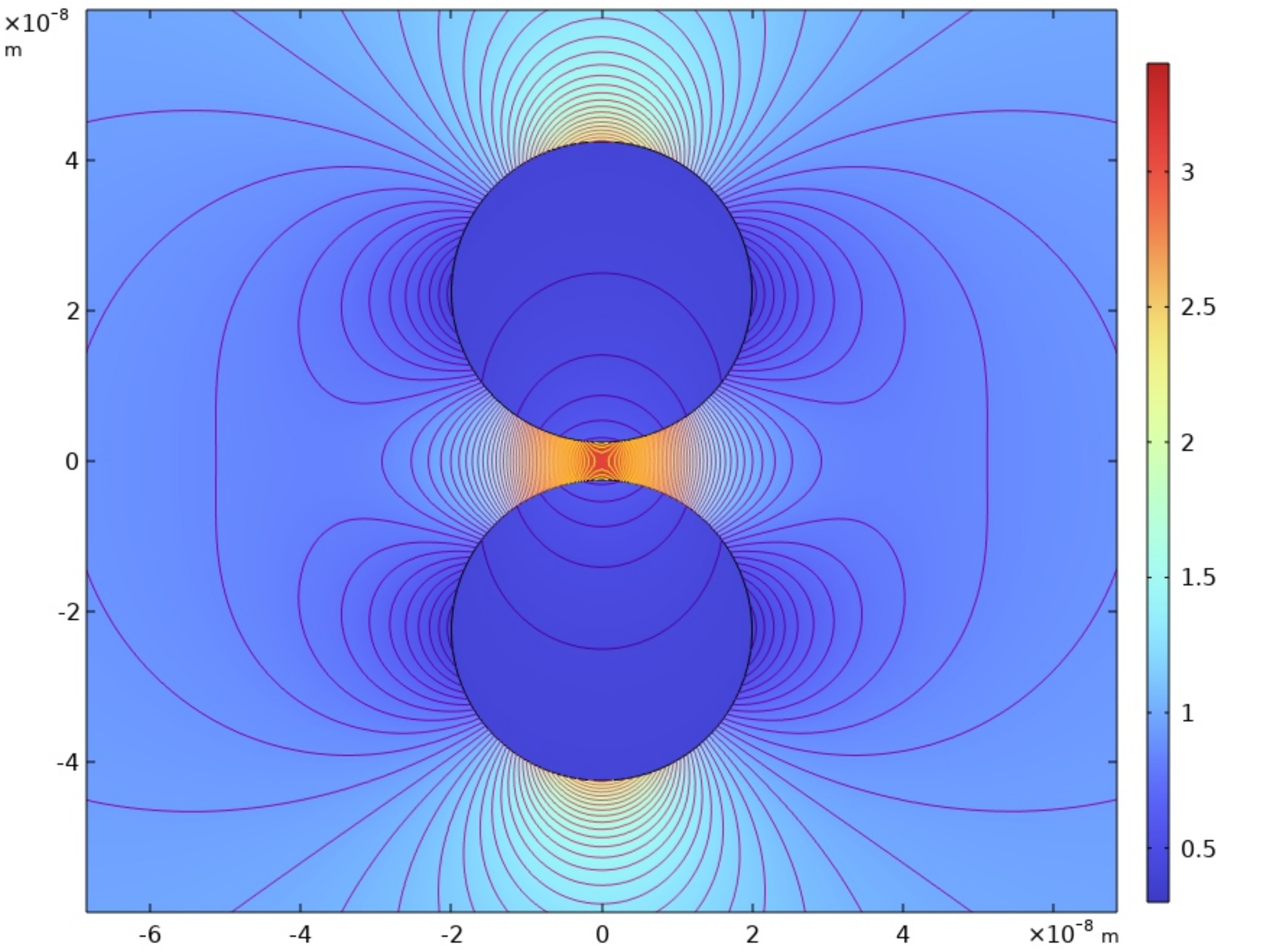}}

\subfigure[]%
{\includegraphics[width=.27\textwidth]{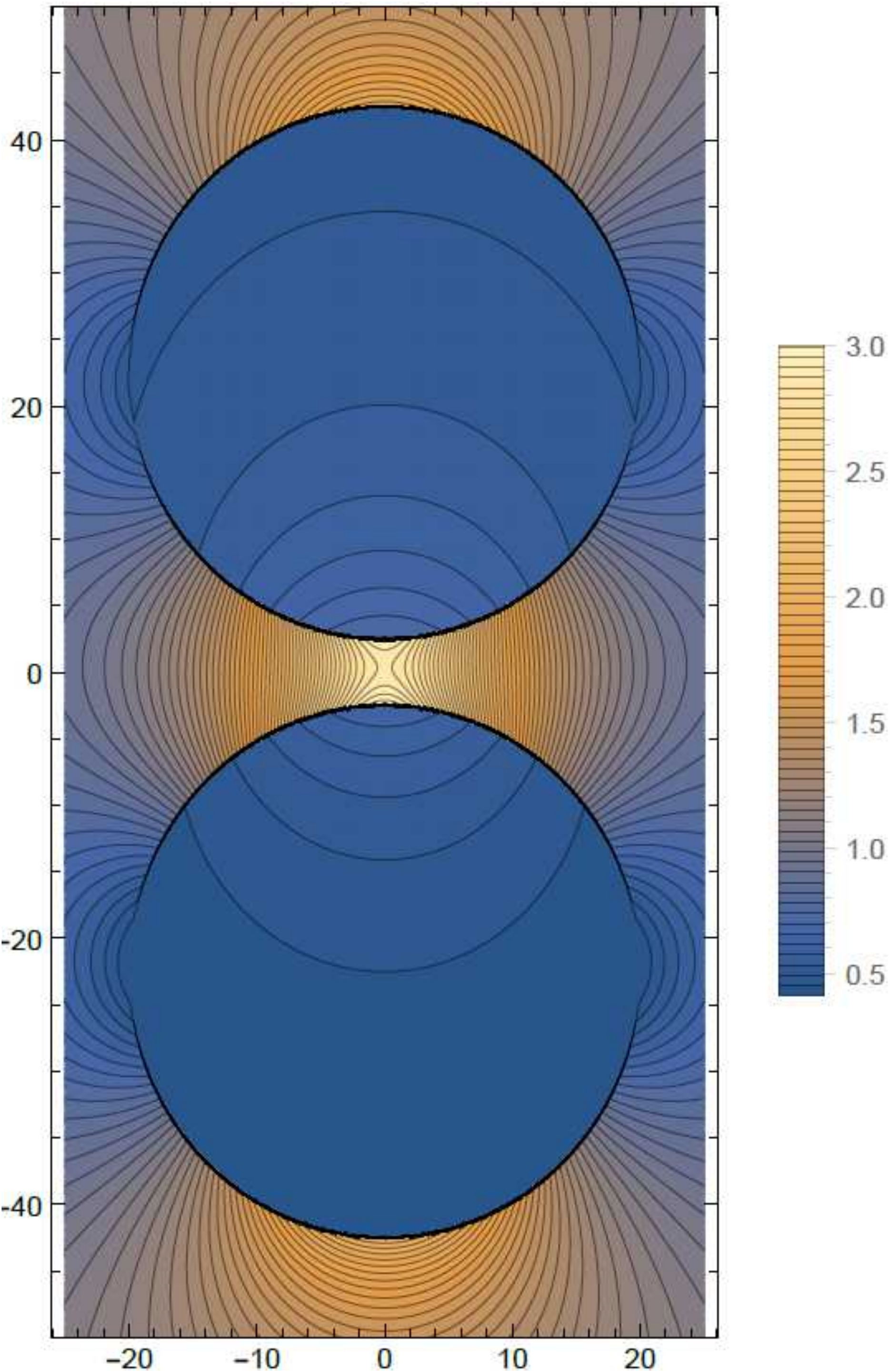}}
\subfigure[]%
{\includegraphics[width=.55\textwidth]{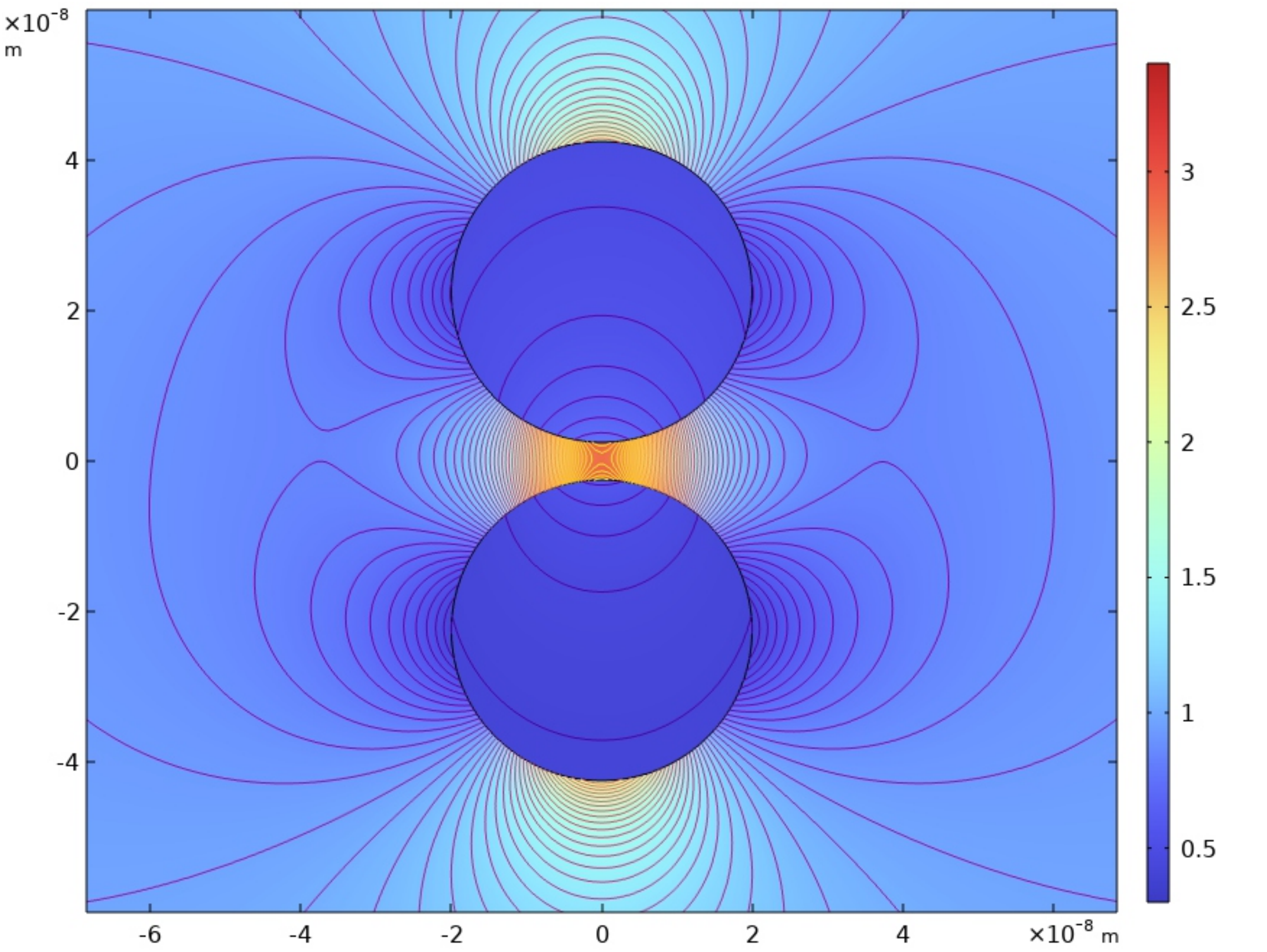}}
\caption{Distribution of the intensity $|E|^{2}$ at $R_{1}=R_{2} = 20$ nm, $\varepsilon_{2} = 2.25$, $\Delta = 5\,$ nm, $\lambda=5\ \mu$m: $\eps_1=2.25$ for (a), (b) and $\eps_1=2$ for (c), (d).}\label{fig:fieldmap}
\end{figure}

\begin{figure}
	\subfigure[]%
	{\includegraphics[width=.27\textwidth]{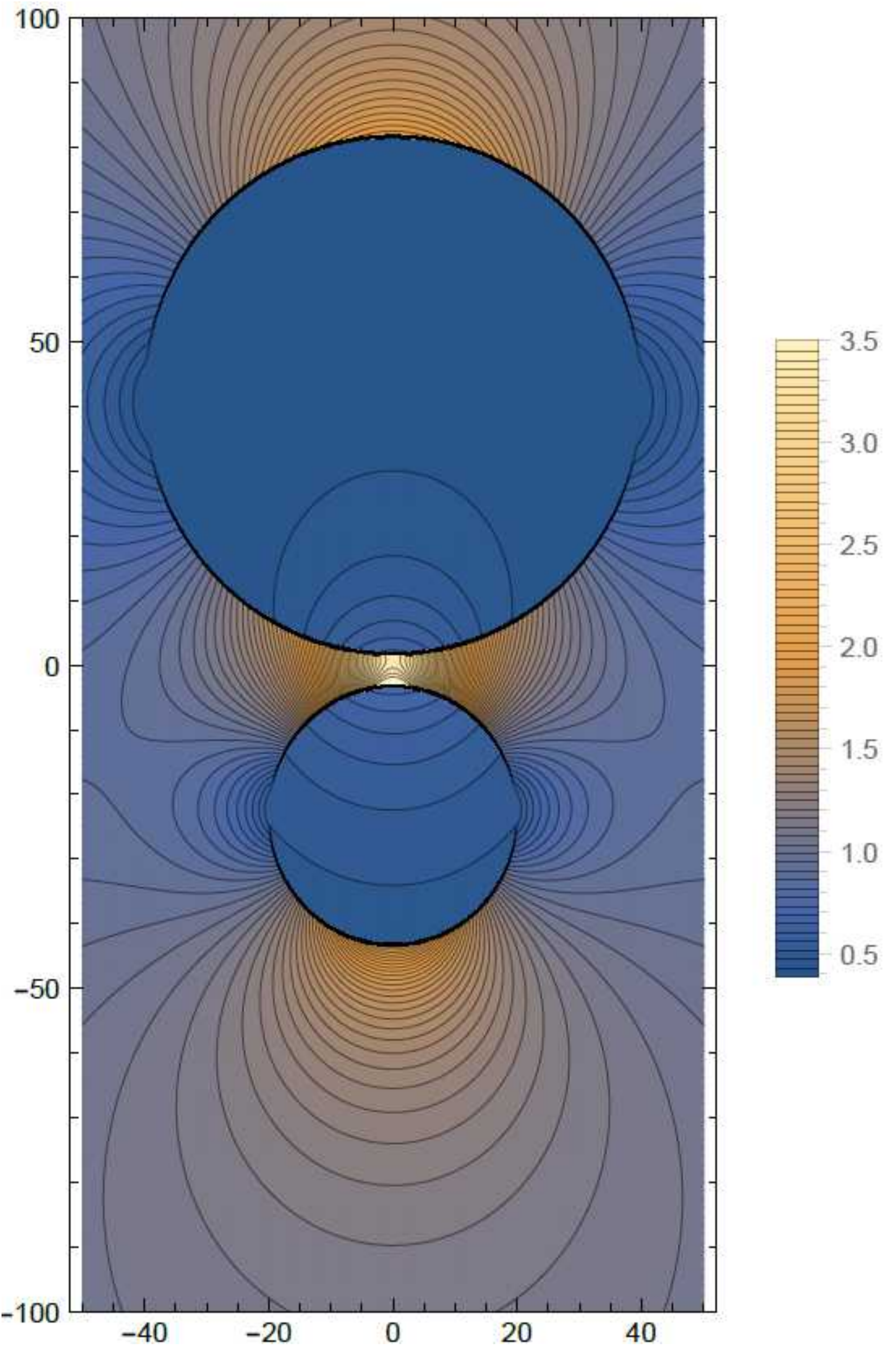}}
	\subfigure[]%
	{\includegraphics[width=.55\textwidth]{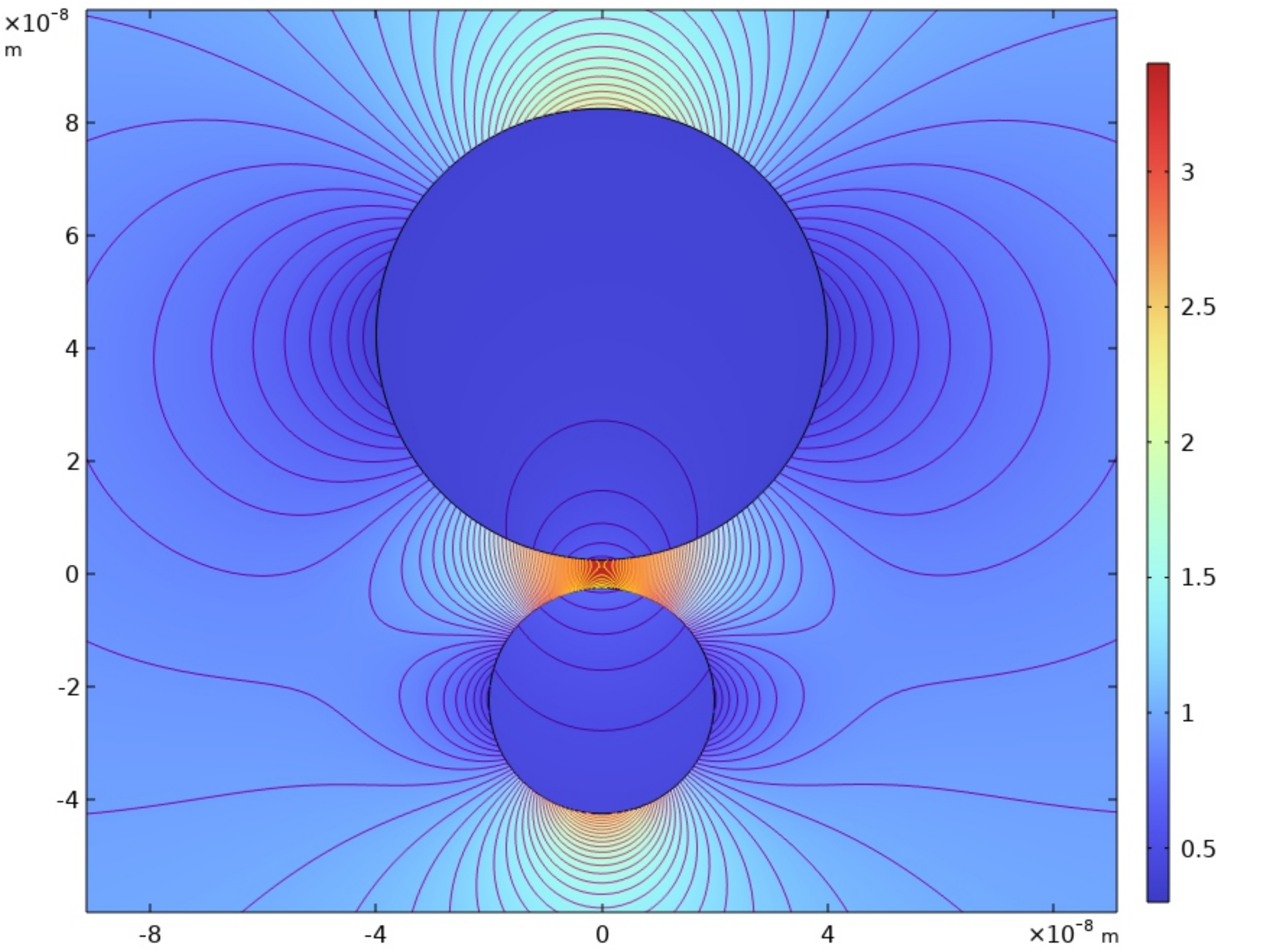}}
	
	\subfigure[]%
	{\includegraphics[width=.27\textwidth]{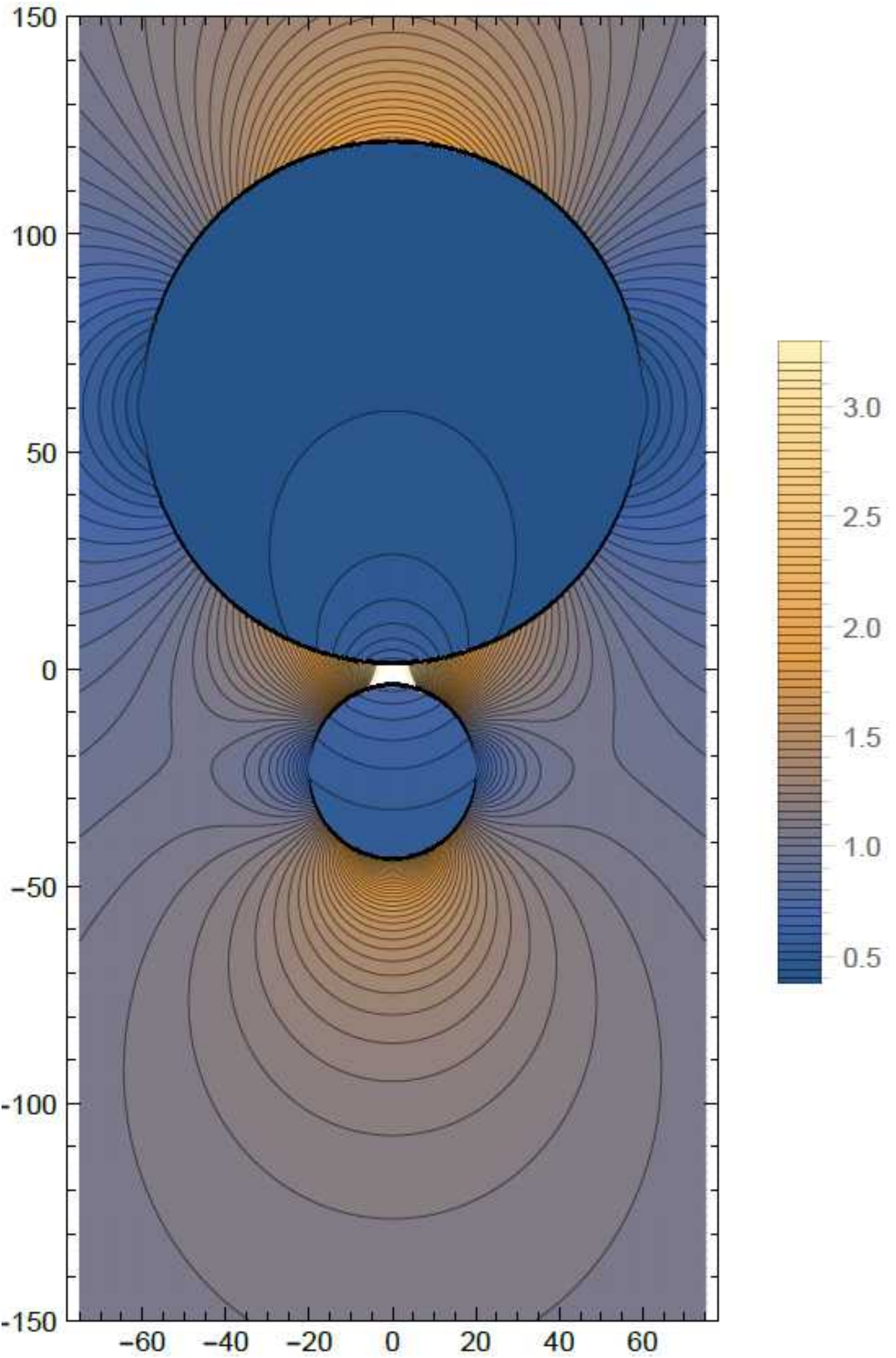}}
	\subfigure[]%
	{\includegraphics[width=.55\textwidth]{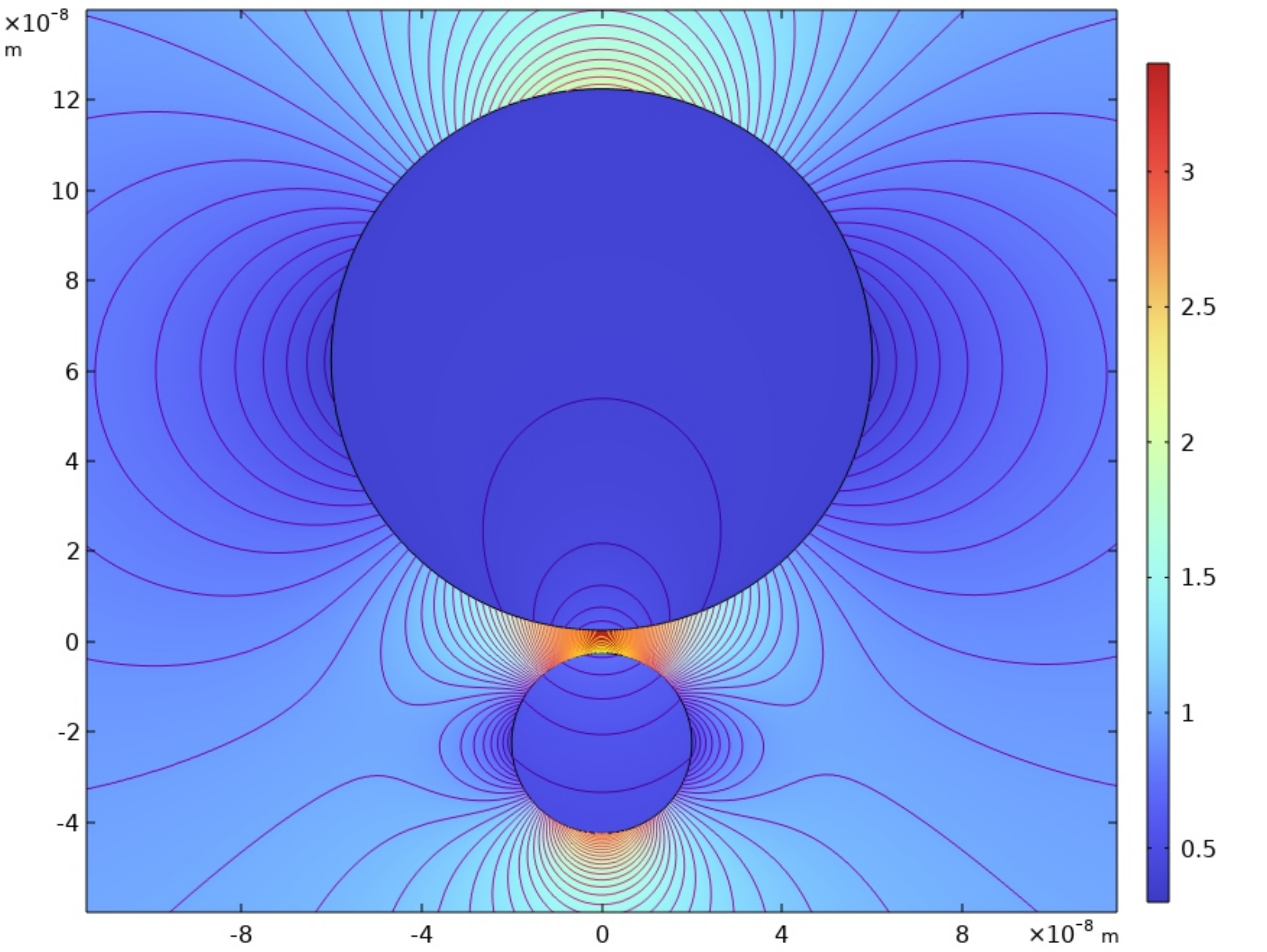}}
	\caption{Distribution of the intensity $|E|^{2}$ at $R_{2} = 20$ nm, $\varepsilon_{1}=\varepsilon_{2} = 2.25$, $\Delta = 5\,$ nm, $\lambda=5\ \mu$m: $R_1=40$ nm for (a), (b) and $R_1=60$ nm for (c), (d).}\label{fig:fieldmapR}
\end{figure}

These formulas are appropriate to find the intensity distribution in the cylinders vicinity. Returning to $(x,y)$ coordinates, we take the gradient to determine the electric field. We keep the first 20 terms from the infinite series as a trade-off between computational complexity and precision. We take the amplitude of incident wave equal to 1. Fig. \ref{fig:fieldmap} and \ref{fig:fieldmapR} present the intensity distribution in arbitrary units. In both figures (a) and (c) are analytical solutions, (b) and (d) are obtained by COMSOL Multiphysics. The level lines demonstrate a good agreement of obtained formulas with the numerical calculation.

We see the expected enhancement of electric field in the point between circles. Local field enhancement by two cylinders due to their interaction and strong optical coupling into a dimer are studied earlier in details \cite{dmitriev2019,bulgakov2019}. For asymmetric situations the point of maximum is shifted with respect to the geometric center.  

\section{Conclusions}\label{sec:conclusions}

We obtain the quasistatic approximation of $p$-wave scattered on two parallel cylinders. For this purpose a conformal transformation in the plane of cross-section is built up that maps the borders into level lines of the new orthogonal coordinate system, the bipolar coordinates. In the symmetric case, where cylinders are of equal diameters and refractive indices, the solution is shown to tend to known formulas. The field can be expressed as a decomposition in terms of the Laplace operator eigenfunctions. The system of algebraic equations for its coefficients has been solved and the near-field intensity distribution has been found. This solution agrees with the numerical calculations by COMSOL Multiphysics. Finally, we build up Green's function, making it possible to calculate the higher orders.

The formulas obtained are useful for further investigation of the scattering process, in particular, for correct calculation of resonant configurations in a nanoparticle structure. The quasistatic approximation is valid while the diameters and gap between the cylinders remain much less than the radiation wavelength.

\section*{Acknowledgements}
The work was supported by Russian Science Foundation, grant \#22-22-00633.  	

\appendix
\setcounter{equation}{0}
\renewcommand{\theequation}{A.\arabic{equation}}
\section*{Appendix: Green's function in bipolar coordinates}
Similar to Eq. (\ref{pot:eq}), we can write the Green function as a series of the eigen functions. In the bipolar coordinates free space is mapped into the strip $\xi_2<\xi<\xi_1$. There are three epressions of the function in different domains:
\setcounter{equation}{0}
\begin{equation}
G(\xi,\xi^{\prime},\eta,\eta^{\prime}) = 
\begin{cases}
G_{1}(\xi,\xi^{\prime},\eta,\eta^{\prime}), \xi^{\prime} > \xi_{1} \\
G_{2}(\xi,\xi^{\prime},\eta,\eta^{\prime}), \xi_{1}>\xi^{\prime} > \xi_{2} \\
G_{3}(\xi,\xi^{\prime},\eta,\eta^{\prime}), \xi_{2} >\xi^{\prime}
\end{cases}
\end{equation}

In the first domain
\begin{equation}
G_{1}(\xi,\xi^{\prime},\eta,\eta^{\prime})=
\begin{cases}
\sum_{m=0}^{+\infty}A^{1}_{m}e^{-m\xi}\cos(m(\eta-\eta^{\prime})), \xi>\xi^{\prime},
\\
\sum_{m=0}^{+\infty}\left(F^{1}_{m}e^{m\xi}+L^{1}_{m}e^{-m\xi}\right)\cos(m(\eta-\eta^{\prime})), \xi^{\prime}>\xi>\xi_{1},
\\
\sum_{m=0}^{+\infty}\left(C^{1}_{m}e^{m\xi}+D^{1}_{m}e^{-m\xi}\right)\cos(m(\eta-\eta^{\prime})), \xi_{1}>\xi>\xi_{2},
\\
\sum_{m=0}^{+\infty}B^{1}_{m}e^{m\xi}\cos(m(\eta-\eta^{\prime})), \xi>\xi^{\prime},
\end{cases}
\end{equation}
where
\begin{eqnarray*}
\begin{gathered}
A^{1}_{n} = -\frac{\alpha e^{-n\xi^{\prime}}\left( e^{4n\xi_{1}}(\varepsilon_{1}-1)(\varepsilon_{2}+1)+
	e^{2n(\xi_{1}+\xi^{\prime})}(\varepsilon_{1}+1)(\varepsilon_{2}+1)\right)}
{2\left(e^{2n\xi_{1}}(\varepsilon_{1}+1)(\varepsilon_{2}+1)-e^{2n\xi_{2}}(\varepsilon_{1}-1)(\varepsilon_{2}-1)\right)}+\\
+\frac{\alpha e^{-n\xi^{\prime}}\left( e^{2n(\xi_{2}+\xi^{\prime})}(\varepsilon_{1}-1)(\varepsilon_{2}-1)+e^{2n(\xi_{1}+\xi_{2})}(\varepsilon_{1}+1)(\varepsilon_{2}-1)\right)}{2\left(e^{2n\xi_{1}}(\varepsilon_{1}+1)(\varepsilon_{2}+1)-e^{2n\xi_{2}}(\varepsilon_{1}-1)(\varepsilon_{2}-1)\right)},
\end{gathered}
\\
B^{1}_{n} = -\frac{2\alpha\varepsilon_{1} e^{2n\xi_{1}-n\xi^{\prime}}}{e^{2n\xi_{1}}(\varepsilon_{1}+1)(\varepsilon_{2}+1)-e^{2n\xi_{2}}(\varepsilon_{1}-1)(\varepsilon_{2}-1)},
\\
C^{1}_{n} = -\frac{\alpha\varepsilon_{1}(\varepsilon_{2}+1) e^{2n\xi_{1}-n\xi^{\prime}}}{e^{2n\xi_{1}}(\varepsilon_{1}+1)(\varepsilon_{2}+1)-e^{2n\xi_{2}}(\varepsilon_{1}-1)(\varepsilon_{2}-1)},
\\
D^{1}_{n} = -\frac{\alpha\varepsilon_{1}(\varepsilon_{2}-1) e^{2n\xi_{1}+2n\xi_{2}-n\xi^{\prime}}}{e^{2n\xi_{1}}(\varepsilon_{1}+1)(\varepsilon_{2}+1)-e^{2n\xi_{2}}(\varepsilon_{1}-1)(\varepsilon_{2}-1)},
\\
L^{1}_{n} = -\frac{\alpha e^{2n\xi_{1}-n\xi^{\prime}}\left(e^{2n\xi_{1}}(\varepsilon_{1}-1)(\varepsilon_{2}+1)-e^{2n\xi_{2}}(\varepsilon_{1}+1)(\varepsilon_{2}-1)\right)}{2\left( e^{2n\xi_{1}}(\varepsilon_{1}+1)(\varepsilon_{2}+1)-e^{2n\xi_{2}}(\varepsilon_{1}-1)(\varepsilon_{2}-1)\right)},
\\
	F^{1}_{n} = -\frac{1}{2}\alpha e^{-n\xi^{\prime}},\quad
\alpha = \frac{C^{2}}{\left(\cosh\xi^{\prime}-\cos\eta^{\prime}
	\right)^{2}}.	
\end{eqnarray*}

The second function is
\begin{equation}
G_{2}(\xi,\xi^{\prime},\eta,\eta^{\prime})=
\begin{cases}
\sum_{m=0}^{+\infty}A^{2}_{m}e^{-m\xi}\cos(m(\eta-\eta^{\prime})), \xi>\xi_{1},
\\
\sum_{m=0}^{+\infty}\left(C^{2}_{m}e^{m\xi}+D^{2}_{m}e^{-m\xi}\right)\cos(m(\eta-\eta^{\prime})), \xi_{1}>\xi>\xi^{\prime},
\\
\sum_{m=0}^{+\infty}\left(F^{2}_{m}e^{m\xi}+L^{2}_{m}e^{-m\xi}\right)\cos(m(\eta-\eta^{\prime})), \xi^{\prime}>\xi>\xi_{2},
\\
\sum_{m=0}^{+\infty}B^{2}_{m}e^{m\xi}\cos(m(\eta-\eta^{\prime})), \xi<\xi_{2},
\end{cases}
\end{equation}
where
\begin{align*}
A^{2}_{n} = -\frac{\alpha e^{2n\xi_{1}-n\xi^{\prime}}\left(-e^{2n\xi_{2}}(\varepsilon_{2}-1)+e^{2n\xi^{\prime}}(\varepsilon_{2}+1)\right)}{ e^{2n\xi_{1}}(\varepsilon_{1}+1)(\varepsilon_{2}+1)-e^{2n\xi_{2}}(\varepsilon_{1}-1)(\varepsilon_{2}-1)},
\\
B^{2}_{n} = \frac{\alpha e^{-n\xi_{3}}\left(e^{2n\xi^{\prime}}(\varepsilon_{1}-1)-e^{2n\xi_{1}}(\varepsilon_{1}+1)\right)}{ e^{2n\xi_{1}}(\varepsilon_{1}+1)(\varepsilon_{2}+1)-e^{2n\xi_{2}}(\varepsilon_{1}-1)(\varepsilon_{2}-1)},
\\
C^{2}_{n} = \frac{\alpha e^{-n\xi_{3}}(\varepsilon_{1}-1)\left(e^{2n\xi^{\prime}}(\varepsilon_{2}+1)-e^{2n\xi_{2}}(\varepsilon_{2}-1)\right)}{ 2\left(e^{2n\xi_{1}}(\varepsilon_{1}+1)(\varepsilon_{2}+1)-e^{2n\xi_{2}}(\varepsilon_{1}-1)(\varepsilon_{2}-1)\right)},
\\
D^{2}_{n} = -\frac{\alpha e^{2n\xi_{1}-n\xi_{3}}(\varepsilon_{1}+1)\left(e^{2n\xi^{\prime}}(\varepsilon_{2}+1)-e^{2n\xi_{2}}(\varepsilon_{2}-1)\right)}{ 2\left(e^{2n\xi_{1}}(\varepsilon_{1}+1)(\varepsilon_{2}+1)-e^{2n\xi_{2}}(\varepsilon_{1}-1)(\varepsilon_{2}-1)\right)},
\\
F^{2}_{n} = -\frac{\alpha e^{-n\xi_{3}}(\varepsilon_{2}+1)\left(e^{2n\xi_{1}}(\varepsilon_{1}+1)-e^{2n\xi^{\prime}}(\varepsilon_{1}-1)\right)}{ 2\left(e^{2n\xi_{1}}(\varepsilon_{1}+1)(\varepsilon_{2}+1)-e^{2n\xi_{2}}(\varepsilon_{1}-1)(\varepsilon_{2}-1)\right)},
\\
L^{2}_{n} = -\frac{\alpha e^{2n\xi_{2}-n\xi_{3}}(\varepsilon_{2}-1)\left(e^{2n\xi^{\prime}}(\varepsilon_{1}-1)-e^{2n\xi_{1}}(\varepsilon_{1}+1)\right)}{ 2\left(e^{2n\xi_{1}}(\varepsilon_{1}+1)(\varepsilon_{2}+1)-e^{2n\xi_{2}}(\varepsilon_{1}-1)(\varepsilon_{2}-1)\right)}.
\end{align*}
	
The third value is
\begin{equation}
G_{3}(\xi,\xi^{\prime},\eta,\eta^{\prime})=
\begin{cases}
\sum_{m=0}^{+\infty}A^{3}_{m}e^{-m\xi}\cos(m(\eta-\eta^{\prime})), \xi>\xi_{1},
\\
\sum_{m=0}^{+\infty}\left(C^{3}_{m}e^{m\xi}+D^{3}_{m}e^{-m\xi}\right)\cos(m(\eta-\eta^{\prime})), \xi_{1}>\xi>\xi_{2},
\\
\sum_{m=0}^{+\infty}\left(F^{3}_{m}e^{m\xi}+L^{3}_{m}e^{-m\xi}\right)\cos(m(\eta-\eta^{\prime})), \xi_{2}>\xi>\xi^{\prime},
\\
\sum_{m=0}^{+\infty}B^{3}_{m}e^{m\xi}\cos(m(\eta-\eta^{\prime})), \xi^{\prime}>\xi,
\end{cases},
\end{equation}
where
\begin{align*}
A^{3}_{n} = -\frac{2\alpha\varepsilon_{2} e^{2n\xi_{1}+n\xi^{\prime}}}{e^{2n\xi_{1}}(\varepsilon_{1}+1)(\varepsilon_{2}+1)-e^{2n\xi_{2}}(\varepsilon_{1}-1)(\varepsilon_{2}-1)},
\\
\begin{gathered}
B^{3}_{n} = -\frac{\alpha e^{-n(2\xi_{2}+\xi^{\prime})}\left( e^{4n\xi_{2}}(\varepsilon_{1}-1)(\varepsilon_{2}-1)+e^{2n(\xi_{2}+\xi^{\prime})}(\varepsilon_{1}-1)(\varepsilon_{2}+1)\right)}{2\left(e^{2n\xi_{1}}(\varepsilon_{1}+1)(\varepsilon_{2}+1)-e^{2n\xi_{2}}(\varepsilon_{1}-1)(\varepsilon_{2}-1)\right)}+\\
+\frac{\alpha e^{-n(2\xi_{2}+\xi^{\prime})}\left( e^{2n(\xi_{1}+\xi^{\prime})}(\varepsilon_{1}+1)(\varepsilon_{2}-1)+e^{2n(\xi_{1}+\xi_{2})}(\varepsilon_{1}+1)(\varepsilon_{2}+1)\right)}{2\left(e^{2n\xi_{1}}(\varepsilon_{1}+1)(\varepsilon_{2}+1)-e^{2n\xi_{2}}(\varepsilon_{1}-1)(\varepsilon_{2}-1)\right)},
\end{gathered}
\\
C^{3}_{n} = \frac{\alpha\varepsilon_{2}(\varepsilon_{1}-1) e^{n\xi^{\prime}}}{e^{2n\xi_{1}}(\varepsilon_{1}+1)(\varepsilon_{2}+1)-e^{2n\xi_{2}}(\varepsilon_{1}-1)(\varepsilon_{2}-1)},
\\
D^{3}_{n} = -\frac{\alpha\varepsilon_{2}(\varepsilon_{1}+1) e^{2n\xi_{1}+n\xi^{\prime}}}{e^{2n\xi_{1}}(\varepsilon_{1}+1)(\varepsilon_{2}+1)-e^{2n\xi_{2}}(\varepsilon_{1}-1)(\varepsilon_{2}-1)},
\\
F^{3}_{n} = -\frac{\alpha e^{-2n\xi_{2}+n\xi^{\prime}}\left(e^{2n\xi_{1}}(\varepsilon_{1}+1)(\varepsilon_{2}-1)-e^{2n\xi_{2}}(\varepsilon_{1}-1)(\varepsilon_{2}+1)\right)}{2\left( e^{2n\xi_{1}}(\varepsilon_{1}+1)(\varepsilon_{2}+1)-e^{2n\xi_{2}}(\varepsilon_{1}-1)(\varepsilon_{2}-1)\right)},
L^{3}_{n} = -\frac{1}{2}\alpha e^{n\xi^{\prime}}.
\end{align*}

\pagebreak

\begin{thebibliography}{10}
	\providecommand{\url}[1]{#1}
	\csname url@samestyle\endcsname
	\providecommand{\newblock}{\relax}
	\providecommand{\bibinfo}[2]{#2}
	\providecommand{\BIBentrySTDinterwordspacing}{\spaceskip=0pt\relax}
	\providecommand{\BIBentryALTinterwordstretchfactor}{4}
	\providecommand{\BIBentryALTinterwordspacing}{\spaceskip=\fontdimen2\font plus
		\BIBentryALTinterwordstretchfactor\fontdimen3\font minus
		\fontdimen4\font\relax}
	\providecommand{\BIBforeignlanguage}[2]{{%
			\expandafter\ifx\csname l@#1\endcsname\relax
			\typeout{** WARNING: IEEEtran.bst: No hyphenation pattern has been}%
			\typeout{** loaded for the language `#1'. Using the pattern for}%
			\typeout{** the default language instead.}%
			\else
			\language=\csname l@#1\endcsname
			\fi
			#2}}
	\providecommand{\BIBdecl}{\relax}
	\BIBdecl
	
	\bibitem{girard2005}
	C.~Girard, ``Near fields in nanostructures,'' \emph{Reports on progress in
		physics}, vol.~68, no.~8, p. 1883, 2005.
	
	\bibitem{shalaev2006}
	V.~M. Shalaev and S.~Kawata, \emph{Nanophotonics with surface plasmons}.\hskip
	1em plus 0.5em minus 0.4em\relax Elsevier, 2006.
	
	\bibitem{stockman2011}
	M.~I. Stockman, ``Nanoplasmonics: The physics behind the applications,''
	\emph{Phys. Today}, vol.~64, no.~2, pp. 39--44, 2011.
	
	\bibitem{novotny2012}
	L.~Novotny and B.~Hecht, \emph{Principles of nano-optics}.\hskip 1em plus 0.5em
	minus 0.4em\relax Cambridge university press, 2012.
	
	\bibitem{devilez2015}
	A.~Devilez, X.~Zambrana-Puyalto, B.~Stout, and N.~Bonod, ``Mimicking localized
	surface plasmons with dielectric particles,'' \emph{Physical Review B},
	vol.~92, no.~24, p. 241412, 2015.
	
	\bibitem{krasnok2018}
	A.~Krasnok, M.~Caldarola, N.~Bonod, and A.~Al{\'u}, ``Spectroscopy and
	biosensing with optically resonant dielectric nanostructures,''
	\emph{Advanced optical materials}, vol.~6, no.~5, p. 1701094, 2018.
	
	\bibitem{enrichi2018}
	F.~Enrichi, A.~Quandt, and G.~C. Righini, ``Plasmonic enhanced solar cells:
	Summary of possible strategies and recent results,'' \emph{Renewable and
		Sustainable Energy Reviews}, vol.~82, pp. 2433--2439, 2018.
	
	\bibitem{homola2008}
	J.~Homola, ``Surface plasmon resonance sensors for detection of chemical and
	biological species,'' \emph{Chemical Reviews}, vol. 108, no.~2, pp. 462--493,
	2008.
	
	\bibitem{shalabney2011}
	A.~Shalabney and I.~Abdulhalim, ``Sensitivity-enhancement methods for surface
	plasmon sensors,'' \emph{Laser \& Photonics Reviews}, vol.~5, no.~4, pp.
	571--606, 2011.
	
	\bibitem{piliarik2012}
	M.~Piliarik, H.~{\v{S}}{\'\i}pov{\'a}, P.~Kvasni{\v{c}}ka, N.~Galler, J.~R.
	Krenn, and J.~Homola, ``High-resolution biosensor based on localized surface
	plasmons,'' \emph{Opt. Express}, vol.~20, no.~1, pp. 672--680, 2012.
	
	\bibitem{xu2019}
	Y.~Xu, P.~Bai, X.~Zhou, Y.~Akimov, C.~E. Png, L.-K. Ang, W.~Knoll, and L.~Wu,
	``Optical refractive index sensors with plasmonic and photonic structures:
	promising and inconvenient truth,'' \emph{Advanced Optical Materials},
	vol.~7, no.~9, p. 1801433, 2019.
	
	\bibitem{bialiayeu2012}
	A.~Bialiayeu, A.~Bottomley, D.~Prezgot, A.~Ianoul, and J.~Albert,
	``Plasmon-enhanced refractometry using silver nanowire coatings on tilted
	fibre bragg gratings,'' \emph{Nanotechnology}, vol.~23, no.~44, p. 444012,
	2012.
	
	\bibitem{arasu2016}
	P.~T. Arasu, A.~S.~M. Noor, A.~A. Shabaneh, M.~H. Yaacob, H.~N. Lim, and M.~A.
	Mahdi, ``Fiber bragg grating assisted surface plasmon resonance sensor with
	graphene oxide sensing layer,'' \emph{Opt. Commun.}, vol. 380, pp. 260--266,
	2016.
	
	\bibitem{yee1966numerical}
	K.~Yee, ``Numerical solution of initial boundary value problems involving
	maxwell's equations in isotropic media,'' \emph{IEEE Transactions on antennas
		and propagation}, vol.~14, no.~3, pp. 302--307, 1966.
	
	\bibitem{purcell73}
	E.~M. Purcell and C.~R. Pennypacker, ``Scattering and absorption of light by
	nonspherical dielectric grains,'' \emph{The Astrophysical Journal}, vol. 186,
	pp. 705--714, 1973.
	
	\bibitem{Draine08}
	B.~T. Draine and P.~J. Flatau, ``Discrete-dipole approximation for periodic
	targets: theory and tests,'' \emph{J. Optical Society of America A}, vol.~25,
	no.~11, pp. 2693--2703, 2008.
	
	\bibitem{brebbia2016boundary}
	C.~A. Brebbia and S.~Walker, \emph{Boundary element techniques in
		engineering}.\hskip 1em plus 0.5em minus 0.4em\relax Elsevier, 2016.
	
	\bibitem{zienkiewicz2005finite}
	O.~C. Zienkiewicz, R.~L. Taylor, and J.~Z. Zhu, \emph{The finite element
		method: its basis and fundamentals}.\hskip 1em plus 0.5em minus 0.4em\relax
	Elsevier, 2005.
	
	\bibitem{VKS7}
	N.~N. Voitovich, B.~Z. Katsenelenbaum, and A.~N. Sivov, ``The generalized
	natural-oscillation method in diffraction theory,'' \emph{Sov. Phys. Usp.},
	vol.~19, no.~9, pp. 337--352, 1976.
	
	\bibitem{vorobev2010}
	P.~Vorobev, ``Electric field enhancement between two parallel cylinders due to
	plasmonic resonance,'' \emph{Journal of Experimental and Theoretical
		Physics}, vol. 110, no.~2, pp. 193--198, 2010.
	
	\bibitem{lei2010}
	D.~Y. Lei, A.~Aubry, S.~A. Maier, and J.~B. Pendry, ``Broadband nano-focusing
	of light using kissing nanowires,'' \emph{New Journal of Physics}, vol.~12,
	no.~9, p. 093030, 2010.
	
	\bibitem{bereza2017}
	A.~Bereza, A.~Nemykin, S.~Perminov, L.~Frumin, and D.~Shapiro, ``Light
	scattering by dielectric bodies in the born approximation,'' \emph{Physical
		Review A}, vol.~95, no.~6, p. 063839, 2017.
	
	\bibitem{bereza2019}
	A.~Bereza, L.~Frumin, A.~Nemykin, S.~Perminov, and D.~Shapiro, ``Perturbation
	series for the scattering of electromagnetic waves by parallel cylinders,''
	\emph{EPL (Europhysics Letters)}, vol. 127, no.~2, p. 20002, 2019.
	
	\bibitem{PhysRevRes2020}
	T.~A. van~der Sijs, O.~El~Gawhary, and H.~P. Urbach, ``Electromagnetic
	scattering beyond the weak regime: Solving the problem of divergent born
	perturbation series by pad\'e approximants,'' \emph{Phys. Rev. Research},
	vol.~2, p. 013308, Mar 2020.
	
	\bibitem{PhysRevA.104.063514}
	J.~A. Rebou\ifmmode~\mbox{\c{c}}\else \c{c}\fi{}as and P.~A. Brand\~ao,
	``Scattering of light by a parity-time-symmetric dipole beyond the first born
	approximation,'' \emph{Phys. Rev. A}, vol. 104, p. 063514, Dec 2021.
	
	\bibitem{Ustimenko2022}
	N.~A. Ustimenko, D.~F. Kornovan, K.~V. Baryshnikova, A.~B. Evlyukhin, and M.~I.
	Petrov, ``Multipole born series approach to light scattering by mie-resonant
	nanoparticle structures,'' \emph{Journal of Optics}, vol.~24, no.~3, p.
	035603, 2022.
	
	\bibitem{belan2015}
	S.~Belan and S.~Vergeles, ``Plasmon mode propagation in array of closely spaced
	metallic cylinders,'' \emph{Optical Materials Express}, vol.~5, no.~1, pp.
	130--141, 2015.
	
	\bibitem{lee2016}
	S.-C. Lee, ``Scattering at oblique incidence by multiple cylinders in front of
	a surface,'' \emph{Journal of Quantitative Spectroscopy and Radiative
		Transfer}, vol. 182, pp. 119--127, 2016.
	
	\bibitem{ablowitz2003}
	M.~J. Ablowitz, A.~S. Fokas, and A.~S. Fokas, \emph{Complex variables:
		introduction and applications}.\hskip 1em plus 0.5em minus 0.4em\relax
	Cambridge University Press, 2003.
	
	\bibitem{brown2009}
	J.~W. Brown and R.~V. Churchill, \emph{Complex variables and
		applications}.\hskip 1em plus 0.5em minus 0.4em\relax McGraw-Hill, 2009.
	
	\bibitem{dmitriev2019}
	A.~A. Dmitriev and M.~V. Rybin, ``Combining isolated scatterers into a dimer by
	strong optical coupling,'' \emph{Physical Review A}, vol.~99, no.~6, p.
	063837, 2019.
	
	\bibitem{bulgakov2019}
	E.~N. Bulgakov, K.~N. Pichugin, and A.~F. Sadreev, ``Evolution of the
	resonances of two parallel dielectric cylinders with distance between them,''
	\emph{Physical Review A}, vol. 100, no.~4, p. 043806, 2019.
	
\end{thebibliography}


\end{document}